"
\documentclass[pre, twocolumn, superscriptaddress]{revtex4-1}

\usepackage{graphicx,subfigure}
\usepackage{dcolumn}
\usepackage{epsfig}
\usepackage{latexsym}
\usepackage{amsfonts}
\usepackage{amssymb}
\usepackage{xcolor}
\usepackage{siunitx}
\usepackage{helvet}
\usepackage{amsmath}

\begin{document}
\title{When bigger is faster: a self-van Hove analysis of the enhanced self-diffusion of non-commensurate guest particles in smectics
	}

\author{M. Paul Lettinga} 
\email{Corresponding autor: p.lettinga@fz-juelich.de}
\affiliation{IBI-4, Forschungszentrum J\"{u}lich, D-52425 J\"{u}lich, Germany}
\affiliation{Laboratory for Soft Matter and Biophysics, KU Leuven, Celestijnenlaan
200D, B-3001 Leuven, Belgium }
\author{Laura Alvarez}
\affiliation{Laboratory for Soft Matter and Biophysics, KU Leuven, Celestijnenlaan
200D, B-3001 Leuven, Belgium }
\affiliation{Centre de Recherche Paul-Pascal, CNRS \& Universit\'e de Bordeaux, 115 Avenue Schweitzer, F-33600 Pessac, France}
\author{Olivera Korculanin} 
\affiliation{IBI-4, Forschungszentrum J\"{u}lich, D-52425 J\"{u}lich, Germany}
\affiliation{Laboratory for Soft Matter and Biophysics, KU Leuven, Celestijnenlaan
200D, B-3001 Leuven, Belgium }
\author{Eric Grelet}
\email{Corresponding autor: eric.grelet@crpp.cnrs.fr}
\affiliation{Centre de Recherche Paul-Pascal, CNRS \& Universit\'e de Bordeaux, 115 Avenue Schweitzer, F-33600 Pessac, France}

\date{\today}
\begin{abstract}
	 
We investigate the anomalous dynamics in smectic phases of short host rods where, counter-intuitively, long guest rod-shaped particles diffusive faster than the short host ones, due to their precise size mismatch. In addition to the previously reported mean-square displacement, we  analyze the time evolution of the Self-van Hove functions $G(r,t)$, as this probability density function uncovers intrinsic heterogeneous dynamics. Through this analysis, we show that the dynamics of the host particles parallel to the director becomes non-gaussian and therefore heterogeneous after the nematic-to-smectic-A phase transition, even though it exhibits a nearly diffusive behavior according to its mean-square displacement. In contrast, the non-commensurate guest particles display Gaussian dynamics of the parallel motion, up to the transition to the smectic-B phase.
Thus, we show that the Self-van Hove function is a very sensitive probe to account for the instantaneous and heterogeneous dynamics of our system, and should be more widely considered as a quantitative and complementary approach of the classical mean-square displacement characterization in diffusion processes.

\end{abstract}


\maketitle

In 1827, the Scottish botanist Robert Brown identified random jittery motion of pollen particles suspended in water through his microscope. Much later, in 1905, Albert Einstein proposed a theory to explain the so-called Brownian motion~\cite{Einstein1905}, which was then experimentally confirmed by the French physicist Jean Perrin in 1916~\cite{Perrin1916}. In Einstein's theory, random diffusion at the colloidal scale is explained by fast thermal fluctuations of the surrounding solvent molecules which continuously collide with the colloidal particles causing a slow diffusion. This separation of time scales leads to a Fickian expression for the mean square displacement (MSD) in $n$ dimensions  $MSD \equiv \langle r^2(t)\rangle = 2nDt$, where $D$ is the colloid translational diffusion coefficient. For a colloidal sphere of diameter $a$, the diffusion through a solvent of viscosity $\eta_0$ is given by  $D_0=\frac{k_bT}{6\pi \eta_0 a}$, i.e. by the ratio between thermal agitation in $k_bT$ and friction. This behavior can be generalized to colloidal particles of any shape and anisometry.
In the case of slender rods of length $L$ and diameter $d$, the diffusion rate along the long axis is twice that of the perpendicular diffusion, $D^0_\|=2D^0_\bot$, with $D^0_\|=\frac{k_B T}{2\pi \eta_0 L}\ln\frac{L}{d}$ \cite{Burgers1938}.
In general, the particle size is directly related to the friction they experience within the solvent, and thus the bigger the particles are, the slower they diffuse. This effect amplifies when the particles are embedded in a crowded host environment such as biological cells, polymer melts, or colloidal crystals~\cite{Banks2005,Dix2008,Sokolov2012,Doi1978,Hermans1982,Weeks2002a,Chaudhuri2007}, which hinders the particle dynamics. In this scenario, large guest particles are slower than small host building blocks ~\cite{Koenderink2003} and small guest particles are faster than large host particles ~\cite{Wong2004,Kang2005,vanderGucht2003}.

In previous research, we have proven an exception to this rule when a lamellar self-organized structure of rods contains longer particles whose size exceeds the typical length scale of the host phase ~\cite{Alvarez2017}, which was recently confirmed by simulations~\cite{Chiappini2020}. Long, non-commensurate, guest particles were shown to be more mobile than the small host particles forming the smectic phase, in contrast to their slower diffusion in the nematic and isotropic liquid phases. To this end, we used two types of filamentous bacteriophages, as they are stiff monodisperse rods of tunable length, which exhibit the full sequence of liquid crystalline mesophases expected for hard rods~\cite{Dogic2006,Grelet2008a,Grelet2014}.
These bacteriophages have been widely used to study the self-diffusion of tracer amounts of labeled rods of the different mesophases, including nematic ~\cite{Lettinga2005a,Modlinska2015}, smectic ~\cite{Lettinga2007,Grelet2008a,Pouget2011,Repula2019} and columnar ~\cite{Naderi2013} phases, for which the mean squared displacements parallel and perpendicular to the rod long axis are accounted by a power law:

\begin{eqnarray}\label{Eq_subdiff}
\langle {r}_{\|,\bot}^2(t) \rangle= 2 n D_{\|,\bot} t^{\gamma_{\|,\bot}}
\end{eqnarray}

\noindent where $D_{\|}$ and $D_{\bot}$ are the parallel and perpendicular self-diffusion coefficients, which are particle concentration dependent, and ${\gamma_{\|,\bot}}$ is the particle diffusivity. The complexity of the diffusion of a system is often expressed by its subdiffusivity with an exponent $\gamma_{\|,\bot}<1$, that is usually found in the most dense phases over a broad time range. 
In a more general perspective, the MSD can also be defined as an ensemble average of

\begin{eqnarray}\label{Eq_r2vanHove}
\langle \mathbf{r}^2(t) \rangle=\int d\mathbf{r} G(\mathbf{r},t)\mathbf{r}^2
\end{eqnarray}

\noindent over the probability density $G(\mathbf{r},t)$ of finding a particle at position $\mathbf{r}$ and at time $t$. When the positions of the particles at each point in time are known, the self-van Hove function (SvH) can be directly obtained from the  histogram of the distribution of $N$ particle positions after a time $t$,

\begin{eqnarray}\label{Eq_vanHove}
G(\mathbf{r},t)=\frac{1}{N}\sum_{i=1}^{N} \delta
[\mathbf{r}+\mathbf{r}_{i}(0)-\mathbf{r}_{i}(t)].
\end{eqnarray}

\noindent Therefore information on the time evolution of the dynamics is partially lost when only using the MSD as large time ranges allowing for accurate fits are necessary for identifying the diffusion regime.

\begin{figure*}
	\begin{center}
		\includegraphics[width=1.6\columnwidth]{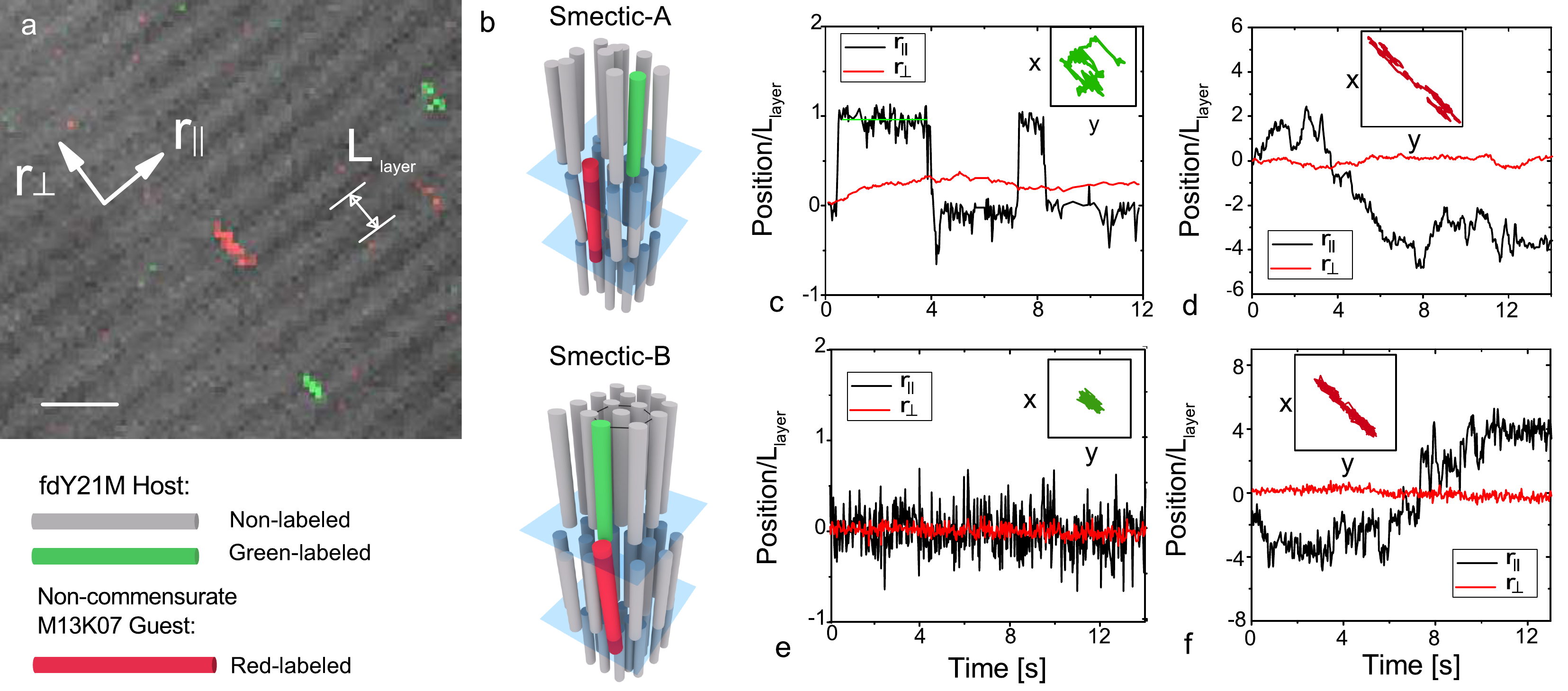}
	\end{center}
	\caption{Guest-host system exhibiting a smectic organization as shown by (a) the overlay of a differential interference contrast (DIC) microscopy image, evidencing smectic layers, and of a fluorescence image, displaying the dual labeling of the host and guest particles. The host smectic phase is formed by short single fdY21M viral rods. A low fraction of fdY21M host particles is labeled with green dyes. Long M13K07 viruses labeled with red fluorescent chromophores are introduced in tracer amount and used as \textit{non-commensurate} guest rods: their length $L_{guest}$ is 1.3 times longer than the typical length scale, $L_{layer}$, associated with the host phase (Smectic layer spacing $L_{layer}\cong L_{host}$). The scale bar represents 2~$\mu m$. (b) Schematic representation of the guest-host system in both smectic-A and smectic-B phases. (c) Example of a single fdY21M host trajectory for which hoping-type events characteristic of a smectic-A phase are observed. (d) Trajectory of a M13K07 guest rod, evidencing its rapid diffusion through the smectic-A lamellar organization. (e) ``Frozen'' dynamics of the host fdY21M particle in smectic-B phase. (f) The non-commensurate-guest particle still exhibits significant motion in the parallel direction when inserted in a smectic-B organization of the host particles.
	}
	
	\label{fig_guesthost}
\end{figure*}

Previously, we merely used the SvH to exemplify the  anomalous hopping-type diffusion of rods  between smectic layers, specifically in the smectic-A phase ~\cite{Lettinga2007,Grelet2008a}, where particles jump by quantized step of one-rod length between adjacent layers. The goal of this paper is to exploit the information contained in the self-van Hove function to highlight the difference in the dynamics between particles that fit within the smectic layers (commensurate hosts) and particles that stick out into both adjacent layers (non-commensurate guests), as depicted in Fig.~\ref{fig_guesthost}. 

The paper is organized as follows.
We will first introduce the function we used to analyse our data and place it in the context of anomalous dynamics.
After a brief experimental section, we will then revisit the MSD data covering the dynamics over the deep nematic range up to the Smectic-A and B phases. Finally, we will discuss the results we have obtained for the self-van Hove functions, showing a distinct behavior not only between the commensurate and non-commensurate particles but also between the different mesophases.

\section{Theoretical background}

When the diffusion is Fickian and isotropic, it follows from the central limit theorem that for sufficiently long times the dynamics, and therefore the SvH, is Gaussian,

\begin{eqnarray}\label{Eq_Gauss}
	G(\mathbf{r},t)\propto\exp\left(-\frac{\mathbf{r}^2}{4Dt} \right)
\end{eqnarray}
 
\noindent In complex fluids, however, this does not generally hold when considering that: 1) these long times are often experimentally inaccessible when particles encounter too many obstacles while diffusing;  2) the Gaussian approximation results from the integration of the Langevin equation under the assumption of spatial isotropy, which is not \textit{a priori} valid for anisotropic complex fluids such as liquid crystals, as recently shown by \citeauthor{Cuetos2018} \cite{Cuetos2018}. Thus, particles can be Brownian while the dynamics is not Gaussian \cite{Wang2012}. Typical examples of non-Gaussian dynamics can be found in colloidal glasses~\cite{Kegel2000,Weeks2000,Weeks2002b}, spheres in random confinement~\cite{Skinner2013}, in entangled filaments~\cite{Wang2012} and in ordered systems~\cite{Wang2009,Turiv2013,Cuetos2018}. A plethora of models for non-Gaussian SvH functions have been suggested ~\cite{Tsallis1995,Wang2012,Chechkin2017}, showing how non-Gaussian dynamics can be explained by assuming a distribution of diffusion rates, caused by a structured host matrix. As this distribution changes in time, there is a ``diffusing diffusivity"~\cite{Chubynsky2014}, resulting in a Laplace distribution~\cite{Chechkin2017},

\begin{eqnarray}\label{Eq_Levi}
G(\mathbf{r},t)\propto \exp \left(-\frac{|\mathbf{r}|}{\sqrt{\langle D \rangle t}}\right) 
\end{eqnarray}
where $\langle D \rangle$ is the averaged diffusion rate which is, in principle, a function of time.
\noindent
As the hoping-type behavior in the smectic phase is related to the availability of free volume in adjacent layers, it is expected that the dynamics is very heterogeneous, depending on the commensurability of the particle size with the energy landscape. We  therefore choose a generalized Gaussian distribution as fitting function for the SvH, 

\begin{eqnarray}\label{Eq_a1a2}
G(r,t)= \frac{\alpha}{\sqrt{4Dt}\Gamma(1/2\alpha)}\exp \left[ - \left(\frac{r^2}{4D t}\right)^{\alpha}\right] 
\end{eqnarray}

\noindent where  $\Gamma$ denotes the gamma function, and 
$r\equiv r_{\|}$ with $\alpha \equiv \alpha_{\|}$ or $r \equiv r_{\bot}$ with $\alpha \equiv \alpha_{\bot}$, depending on the considered direction with respect to the normal of the smectic layers.
This function does not have a clear separation of time scales, as we expect for our system, and continuously connects a Laplacian, where $\alpha_{\|,\bot} \rightarrow 0.5$ and the averaged diffusion rate $\langle D \rangle$ changes with time, with a purely Gaussian dynamics, where $\alpha_{\|,\bot} \rightarrow 1$ and $D\rightarrow constant$. Note that the factor 4 in the denominator of the exponential function is present in both limits, contrary to what has been suggested in Refs. ~\cite{Chubynsky2014,Chechkin2017}.

\section{Experimental section}
As monodisperse colloidal rods, we used the filamentous rod-like viruses. Thanks to biological engineering, the production of viruses of tunable length and stiffness can be achieved. Specifically, two mutants have been chosen to create the guest-host system studied here: fdY21M virus as a short stiff host (contour length  $L_{host} = 0.91~\mu m$, persistence length  $P_{host} = 9.9~\mu m$, diameter  $d = 7$~nm) and  M13K07 helper phage as long guest semi-flexible rod ($L_{guest} = 1.2~\mu m$,  $P_{guest} = 2.8~\mu m$, $d = 7$~nm) ~\cite{Barry2009,Pouget2011,Sharma2014}, both prepared following standard biological protocols ~\cite{Maniatis1986}. Consequently, the guest-host length ratio is non-commensurate, $L_{guest}/L_{host}\cong L_{guest}/L_{layer}= 1.3$, as shown in Fig.~\ref{fig_guesthost}. 
FdY21M and M13K07 batches were labeled with green (Alexa488-TFP, Invitrogen)  and red (Dylight549-NHS Ester, ThermoFisher) fluorescent dyes, respectively. Labeled particles were added in a ratio of one labeled particle over $10^5$ non-labeled particles such that trajectories of individual rods can be recorded (Fig. \ref{fig_guesthost}). A set of samples with concentrations in the range from the nematic to the smectic phase were prepared (in TRIS-HCl-NaCl buffer, pH 8.2, ionic strength of 20~mM), and single particle tracking was performed using a fluorescence microscope (IX-71 Olympus), equipped with a high-numerical aperture (NA) oil objective (100x PlanApo NA 1.40) and an excitation light source (X-cite series 120 Q). A dual emission image splitter (Optosplit II Andor) was used to simultaneously acquire the two fluorescent emission wavelengths on the sensor of an ultra-fast electron-multiplying camera (NEO sCMOS Andor). A few hundreds of trajectories per concentration were collected using a particle tracking algorithm developed with MATLAB (MathWorks).

\section{Results}

\subsection{Qualitative dynamics}
Figure \ref{fig_guesthost} shows examples of trajectories recorded in the smectic-A phase ($C_{host}= 91$~mg/ml) and smectic-B ($C_{host}= 98$~mg/ml)
for M13K07 guests and fdY21M hosts. In the smectic-A phase, the trajectories of the host particles display discrete steps in the direction of the particle long axis, consistent with earlier observations ~\cite{Lettinga2007,Pouget2011}. Contrary to commensurate host particles, non-commensurate M13K07 guest particles do not exhibit clear hopping-type events. Rather, they exhibit larger parallel displacement reminiscent of the nematic motion along the director $r_{\|}$ (normal of the smectic layers), while the perpendicular displacement $r_{\perp}$ is similar to the one of the commensurate hosts \cite{Alvarez2017}.


In addition, Fig.~\ref{fig_guesthost}e shows that the host dynamics in the smectic-B is highly constrained to in-layer diffusion, with an 
absence of jumping events due to the crystalline order. 
Although the non-commensurate guest particles still exhibit smooth parallel motion along the host layers, their displacement is reduced both by the increase of the host packing fraction and the higher smectic ordering potential \cite{Alvarez2017}.  

As standard and usual characterization of the dynamics, we will first discuss the self-diffusion of both particles in terms of MSD for the parallel and perpendicular directions over a broad range of host concentrations, before exploring the dynamical insights obtained by analyzing quantitatively the SvH. 

\begin{figure}
	\begin{center}
		\includegraphics[width=\columnwidth]{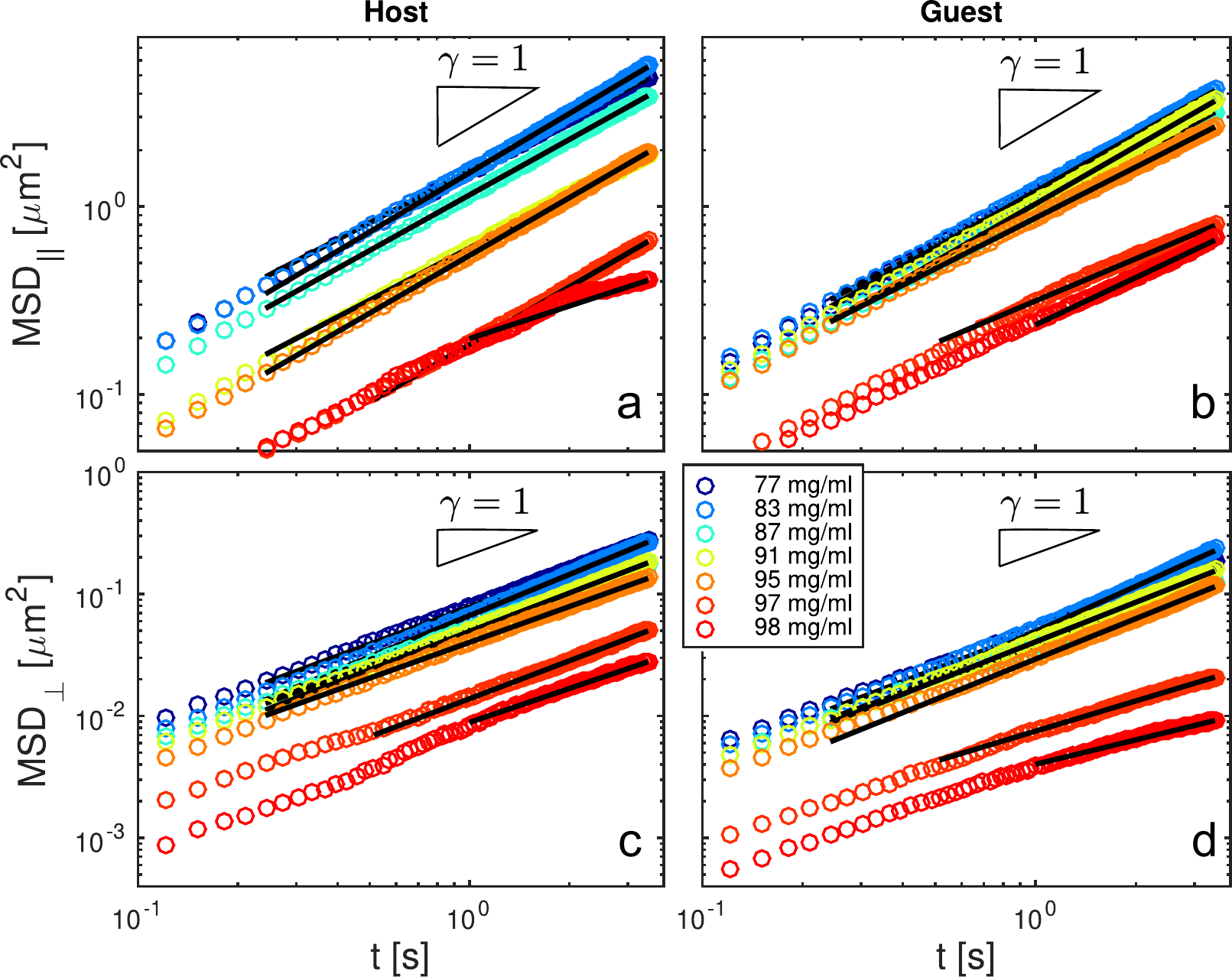}
	\end{center}		
		\caption {Mean square displacement (MSD) parallel (a,b) and perpendicular (c,d) to the director as a function of time for host (a,c) and guest (b,d) particles over a range of concentrations from deep nematic to smectic-B phase. The black lines are power law fits according to Eq. \ref{Eq_subdiff}.
		}
		\label{Fig_msd}
\end{figure}

\begin{figure} 
	\begin{center}
		\includegraphics[width=\columnwidth]{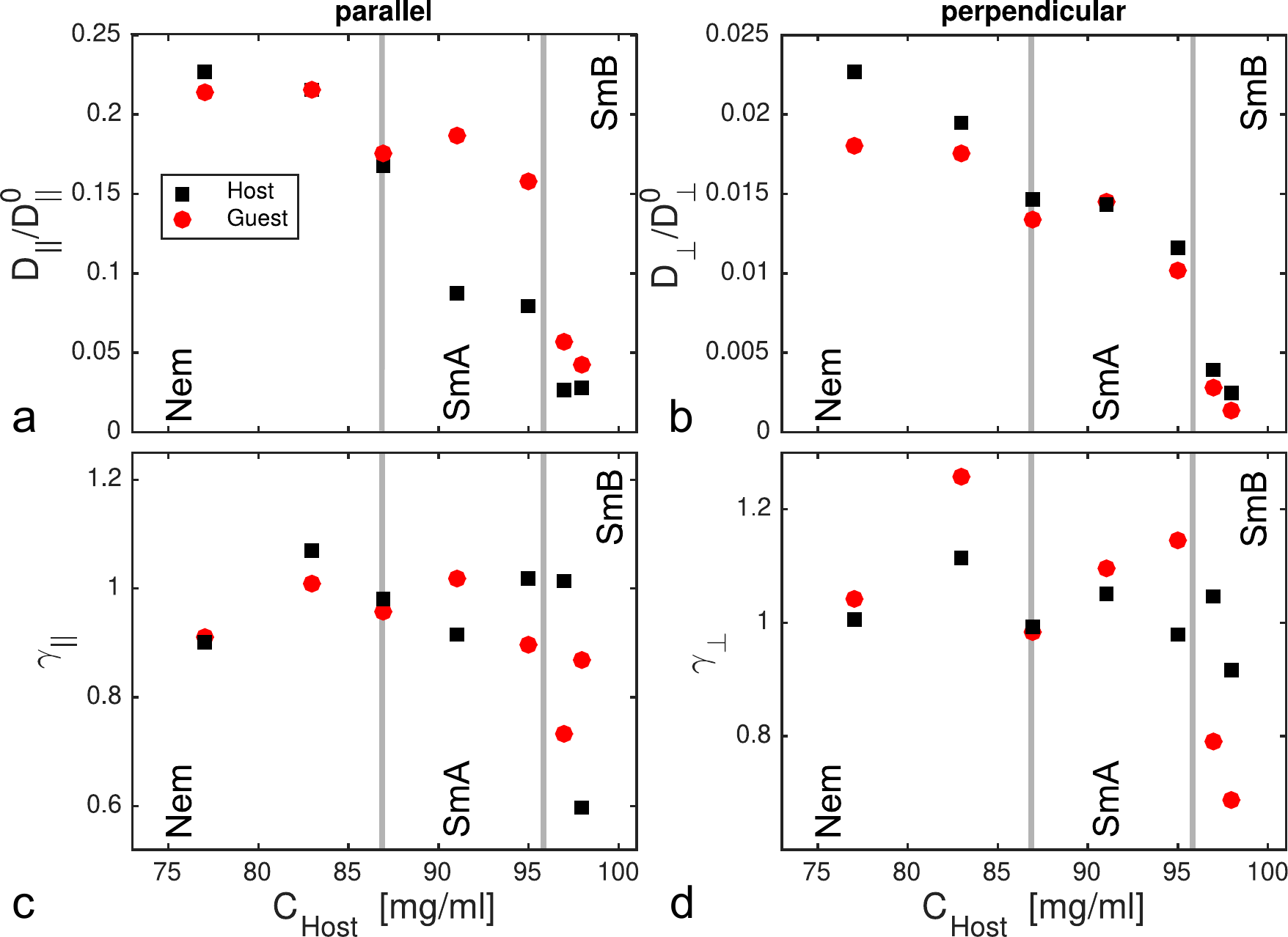}
	\end{center}
		\caption{Dynamics at high concentrations steming from MSD data as represented in Fig. \ref{Fig_msd}. The values of diffusion coefficients (a,b) and diffusion exponents (c,d) are obtained by fitting the MSD with Eq. \ref{Eq_subdiff} for guest (red) and host (black) rods. The diffusion coefficients are scaled by the the ones at infinite dilution, $D^0_{\|,\bot}$. The gray lines indicate the phase transitions. }
			\label{Fig_mdsres}
\end{figure}

\subsection{Mean Square Displacements}

The MSD for the diffusion parallel and perpendicular to the long axis of the guest and host particles  are plotted in Fig.~\ref{Fig_msd} for a wide concentration range including the three liquid crystalline phases.
The diffusion coefficients $D_{\|}$ and $D_\bot$ and the corresponding exponents $\gamma_{\|,\bot}$ have been determined by fitting the data with Eq. \ref{Eq_subdiff}, focusing on the long time range where the rods will have probed the full ordering potential in the smectic phase.

The results of the MSD fits for both guest and host particles are plotted in Fig. \ref{Fig_mdsres}, normalized by the diffusion rates at infinite dilution, $D^0_\|$ and $D^0_\bot$, as introduced above, to account for the trivial rod size dependence of the dynamics.
We first compare the scaled \textit{parallel} diffusion rates, $D_\|/D^0_\|$, as shown in Fig. \ref{Fig_mdsres}a. 
The normalized diffusion rates  in the nematic phase of both long and short rods remarkably overlap within the error bar of their determination. This means that the length of the guest rods does not affect their diffusion rate, given the nematic ordering in the system by the short host.

After the N-SmA transition, the diffusion rates of the host particles $D_{\|}^{host}$ decreases up to the point where the host rods are almost completely immobilized when reaching the crystalline smectic-B phase. This is in strong contrast to the non-commensurate long guests, where $D_{\|}^{guest}$ seems unaffected by the N-SmA phase transition, showing that the long non-commensurate guest particles diffuse significantly \textit{faster} in the smectic-A in contrast to the short host ones. The exponent for both the commensurate and non-commensurate rods remains close to 1 even when the diffusion rate has collapsed, see Fig. \ref{Fig_mdsres}c. This suggests that the motion is diffusive \textit{at long time} 
up to the point that the smectic-B phase is reached.

\begin{figure} 
	\begin{center}
		\includegraphics[width=0.55\columnwidth]{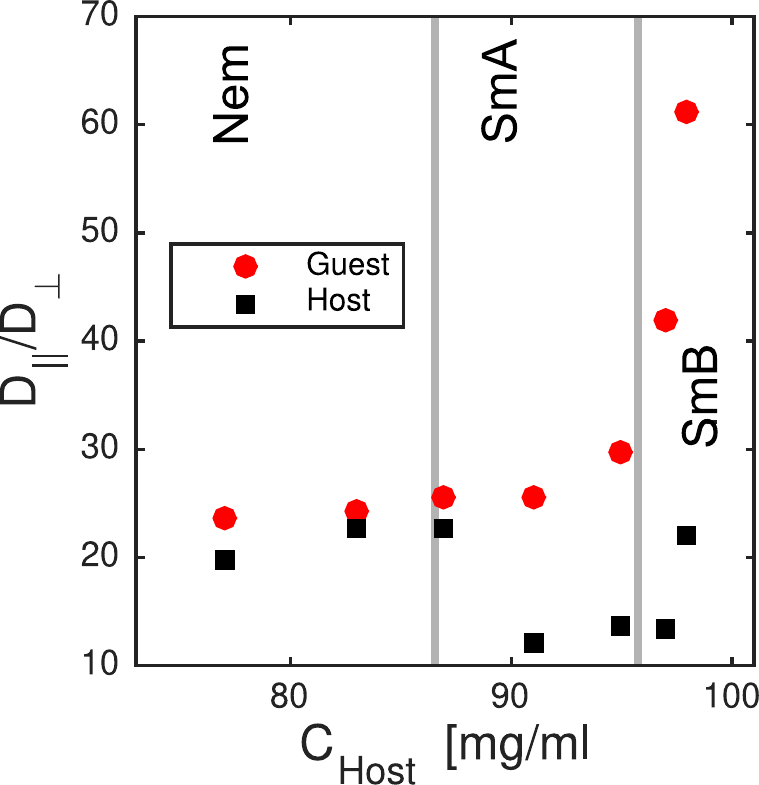}
	\end{center}
		\caption{Anisotropy in the diffusion presented as the ratio of the parallel ($D_\|$) over the perpendicular ($D_\bot$) diffusion corefficients for guest (red) and host (black) particles. The gray lines indicate the phase boundaries between the different liquid crystalline phases.}
		\label{Fig_diffani}
\end{figure}

The \textit{perpendicular} diffusion rate in the nematic phase of the long guests $D_{\bot}^{guest}$ is significantly slower than the  perpendicular diffusion $D_{\bot}^{host}$ of the hosts, as can be seen in Fig. \ref{Fig_mdsres}b.
This can be  understood in terms of the number of encounters a rod will have when moving in both directions. When moving along the long axis, this number will be the same as both rods have exactly the same projection in this direction, given by the diameter of the rod. The diffusion along the long axis should thus not be affected by the length, when scaled by the diffusion at infinite dilution. When a rod is moving in the direction perpendicular to its long axis, then the number of encounters increases linearly with its length.
This effect is partly compensated in the smectic-A phase as most perpendicular diffusion is effectively coupled to the parallel diffusion thanks to jump events, which is for the guest rod faster in the smectic-A phase than for the host particles, so that here $D_{\bot}^{guest}\approx D_{\bot}^{host}$. This perpendicular diffusion seems unaffected both for guests and hosts, up to the point where the smectic-B is entered, at 97 $mg/ml$, after which it strongly decreases. 
As a result, one of the most sensitive parameters to quantify the different dynamic behavior between commensurate and non-commensurate rods is the ratio of the parallel and perpendicular diffusion rates, as plotted in Fig. \ref{Fig_diffani}. 
It shows that the trend of an increasing ratio with increasing ordering in the nematic phase continues for the guest particles into both smectic-A and B phases, in strong contrast to the host viruses, for which this ratio decreases. 
The very high ratio in the smectic-B for the guest is due to the almost complete immobilization of perpendicular diffusion, while it still creates space to move in the parallel direction into the adjacent layers.

\begin{figure*} 
	\begin{center}
		\includegraphics[width=1.5\columnwidth]{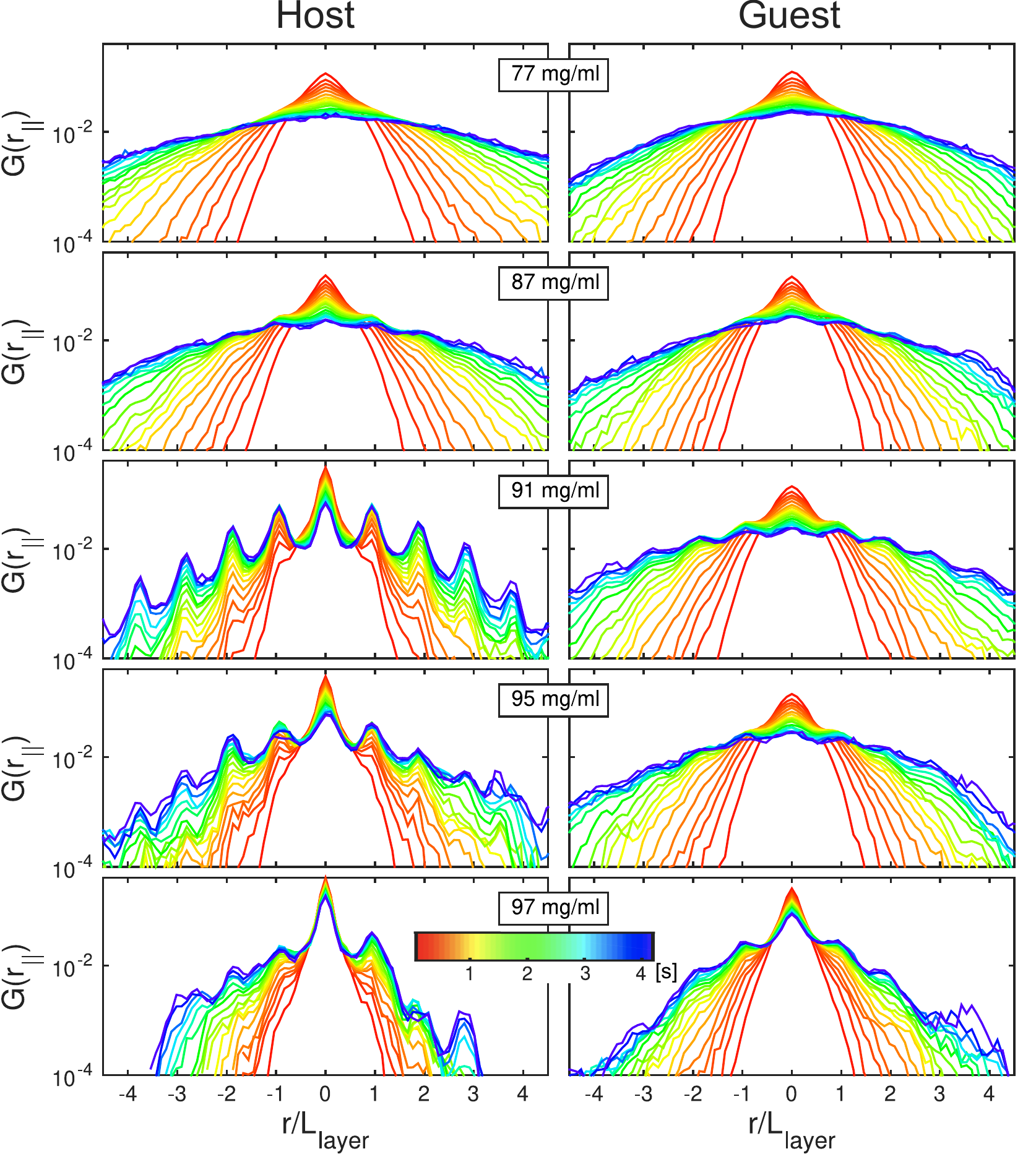}
		\caption{Self-van Hove functions $G(r_{\|},t)$  at increasing times, for the host (left) and guest (right) particles along the normal of the smectic layer. The functions are normalized to one and the positions are renormalized by the smectic layer spacing $L_{layer}$. 
		}
		\label{fig_selfvanhove_para}
	\end{center}
\end{figure*}

\begin{figure*} 
	\begin{center}
		\includegraphics[width=1.5\columnwidth]{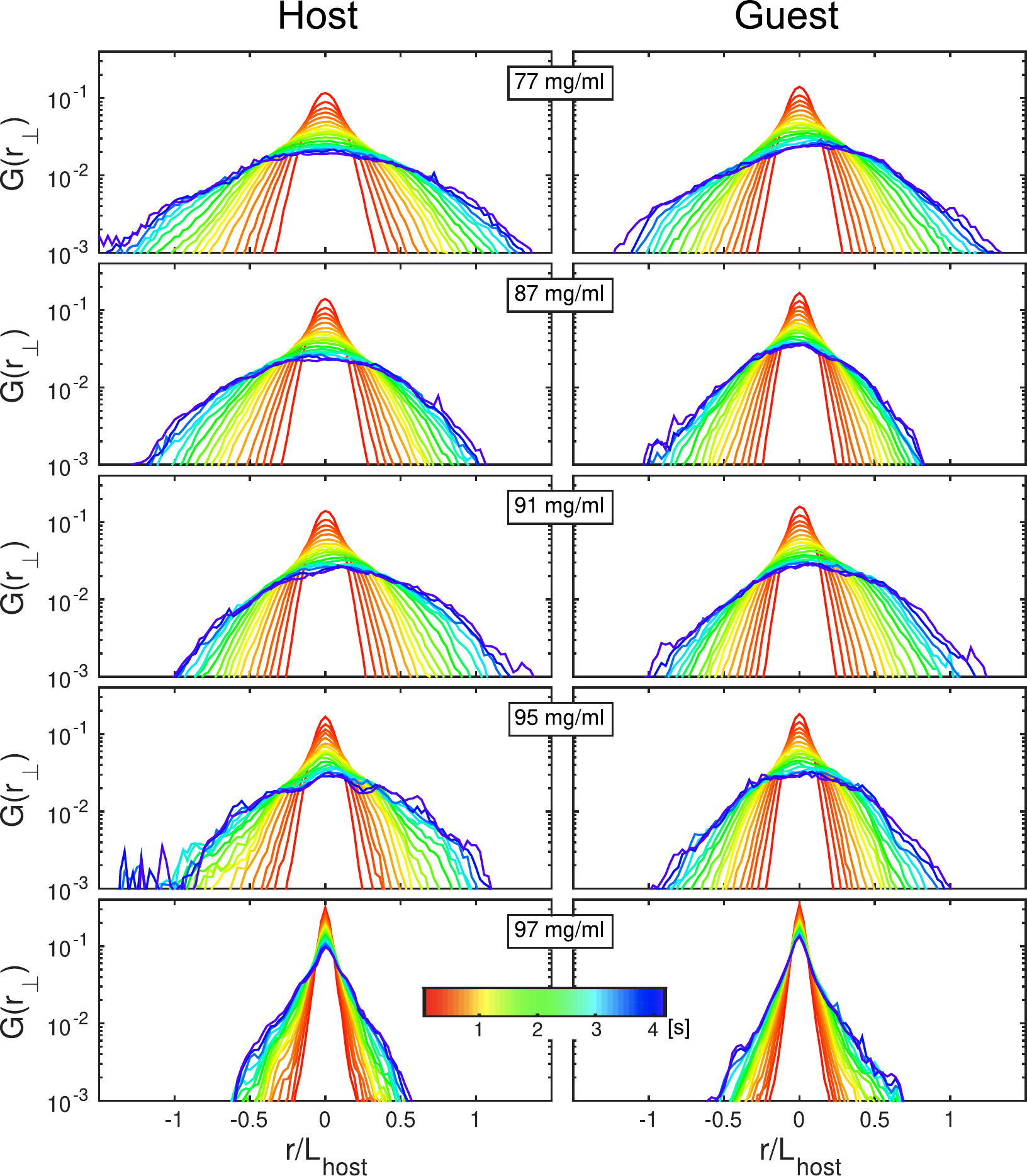}
		\caption{Self-van Hove functions $G(r_{\perp},t)$  at increasing times for the guest (left) and host (right) particles perpendicular to the normal of the smectic layer. The functions are normalized to one and the positions are renormalized by the smectic layer spacing $L_{layer}$.  
		}
		\label{fig_selfvanhove_perp}
	\end{center}
\end{figure*}

\begin{figure} 
	\begin{center}
		\includegraphics[width=0.95\columnwidth]{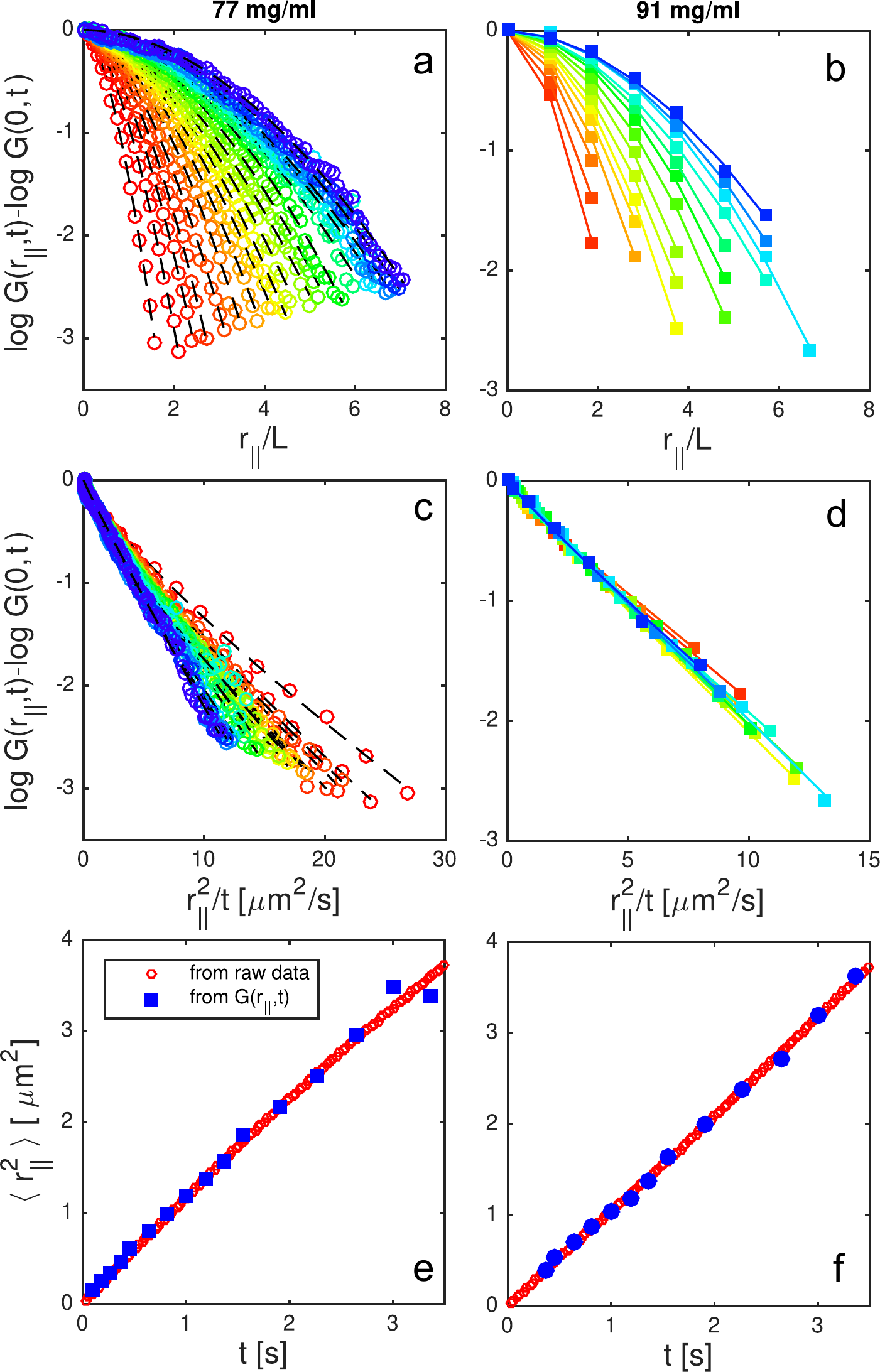}
	\end{center}
		\caption{Self-van Hove functions and the corresponding fits (lines) according to $\log G(r_{\|},t)-\log G(0,t) = -(4\langle D_{\|}\rangle t)^{-\alpha_{\|}}r^{2\alpha_{\|}}/\ln 10$ for the host particles
		at a concentration of (a) 77 mg/ml using the full curve and (b) 91 mg/ml, using only the values at integer numbers of $L_{layer}$, at increasing times. The color coding is the same as in Figs. \ref{fig_selfvanhove_para} and \ref{fig_selfvanhove_perp}. (c,d): as (a,b) but plotting $\log G(r_{\|},t)-\log G(0,t)$ vs  $r_{\|}^2/t$  to highlight the deviations in the diffusivity, as given by the slope of these curves in this representation. (e,f): MSD calculated after integration from the resulting self-can Hove functions (blue bullets) and directly from the measured positions (red circles) showing the self-consistency of our approach. }
		\label{scaledSvH}
\end{figure}

\subsection{Self-van Hove analysis}

The self-van Hove functions underlying the MSDs are reported in Fig. \ref{fig_selfvanhove_para} and \ref{fig_selfvanhove_perp} for the parallel and  perpendicular diffusion, respectively.
In the nematic phase ($C_{host}< 87$~mg/ml),  $G(r_{\|},t)$ is a smooth distribution that smears out over time as expected for Brownian particles. In the smectic-A range, the self-van-Hove functions of the host particles exhibit distinct peaks at integer multiples of the smectic layer spacing of the host phase that accounts for the hopping-type diffusion by indicating an increase of the probability of presence within the layers.
For guest particles at the same concentrations, $G(r_{\|},t)$ are in a first approximation monotonic with more extended ``wings", revealing a higher probability of larger displacements along the parallel direction. Furthermore, the very shallow peaks in $G(r_{\|},t)$ confirms that the guest non-commensurate viruses do not primarily feel the effect of the underlying smectic ordering potential as strong as the host commensurate particles do. This behavior is observed up to the smectic-B phase ($C_{host}\approx 97$~mg/ml in Fig. \ref{fig_selfvanhove_para}), for which some displacement of the guest particles can still be observed, while the diffusion is mostly frozen for the short host rods. 

To identify the dynamics at hand, we fit $\ln G(r,t)-\ln G(0,t) = -(4 \langle D \rangle t)^{-\alpha}r^{2\alpha}$, see Eq. \ref{Eq_a1a2}, providing $\langle D \rangle$ and $\alpha$. For the curves where we observe distinct peaks in the distribution, as for the smectic-A phase of the hosts, we  fit the envelope of the distribution, which are the values at integer numbers of $L_{layer}$ where the probability of finding a particle reaches a local maximum. This means that we  exclude the low probabilities in between the peak positions at $L_{layer}$. Moreover, particles need to have a finite probability for diffusing at least two rod lengths in the parallel direction in order to have enough data points for the fitting. For this reason, numerical fits can only be performed at long times for high concentrations, and no fit at all is possible for host particles above 97 mg/ml and for guest particles above 98 mg/ml.
As can be seen in Fig. \ref{scaledSvH}a, the functional description of the SvH with Eq. \ref{Eq_a1a2} is satisfactory in the nematic as well as in the smectic phase (Fig. \ref{scaledSvH}b) for all probed times. A cross-check has been performed by comparing the MSD calculated directly from the data with the MSD as calculated from the time dependent function after integration according to Eq. \ref{Eq_r2vanHove} (Fig. \ref{scaledSvH}e and f). Both are found in good agreement with each others, even  though the SvH results are somewhat more scattered at long times due to the decreasing statistics in our experimental particle tracking.  

\begin{figure} 
	\begin{center}
		\includegraphics[width=0.95\columnwidth]{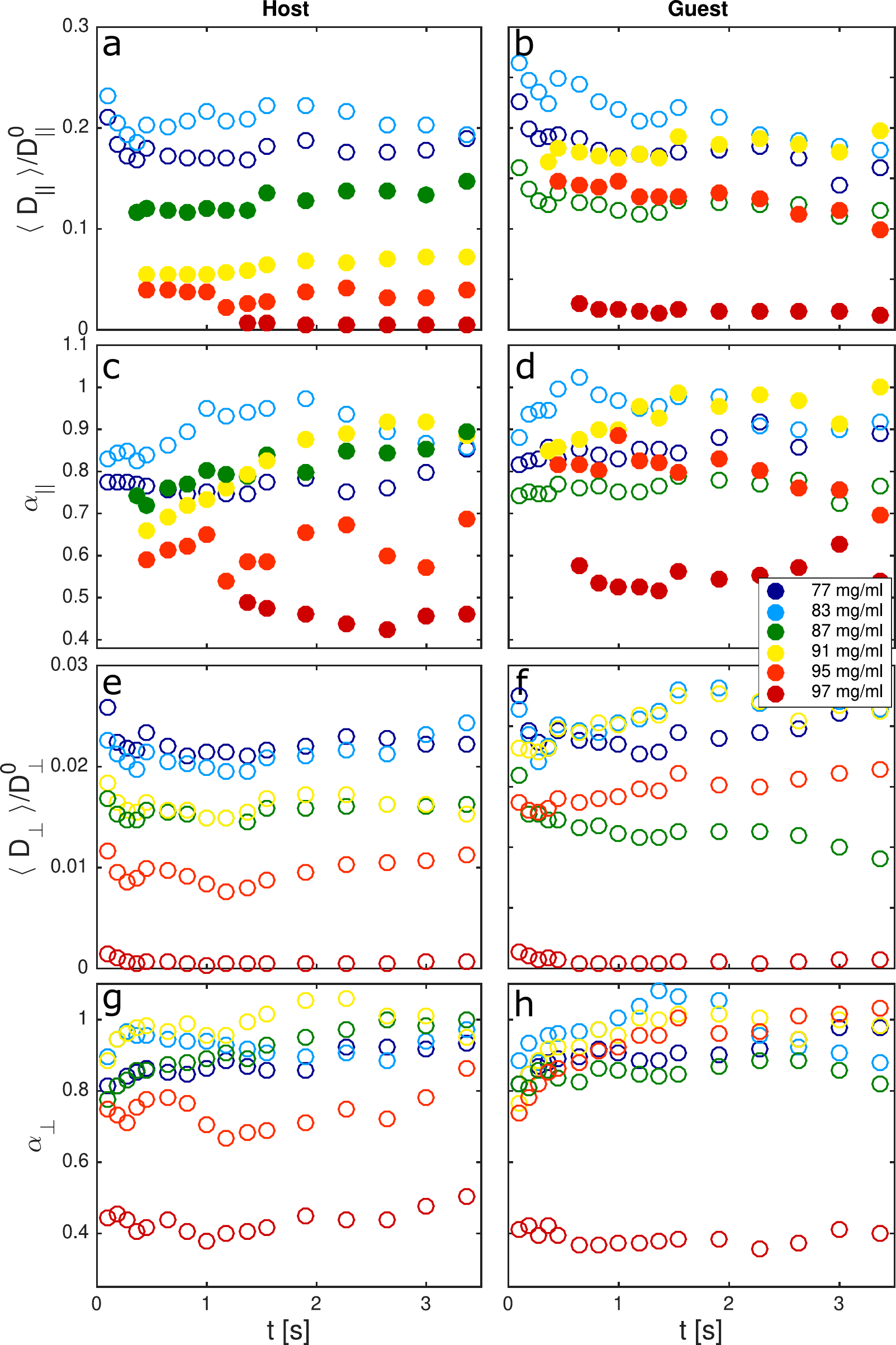}
	\end{center}
		\caption{Time dependence of $\langle D_{\|,\bot} \rangle$ (a,b,e,f) and $\alpha_{\|,\bot}$ (c,d,g,h) for host (a,c,e,g) and guest (b,d,f,h) particles. The solid symbols in (a-d) indicate results obtained by fitting the envelope of the SvH at the peak values.}
		\label{Fig_svh_results}
\end{figure}

\begin{figure} 
	\begin{center}
		\includegraphics[width=0.95\columnwidth]{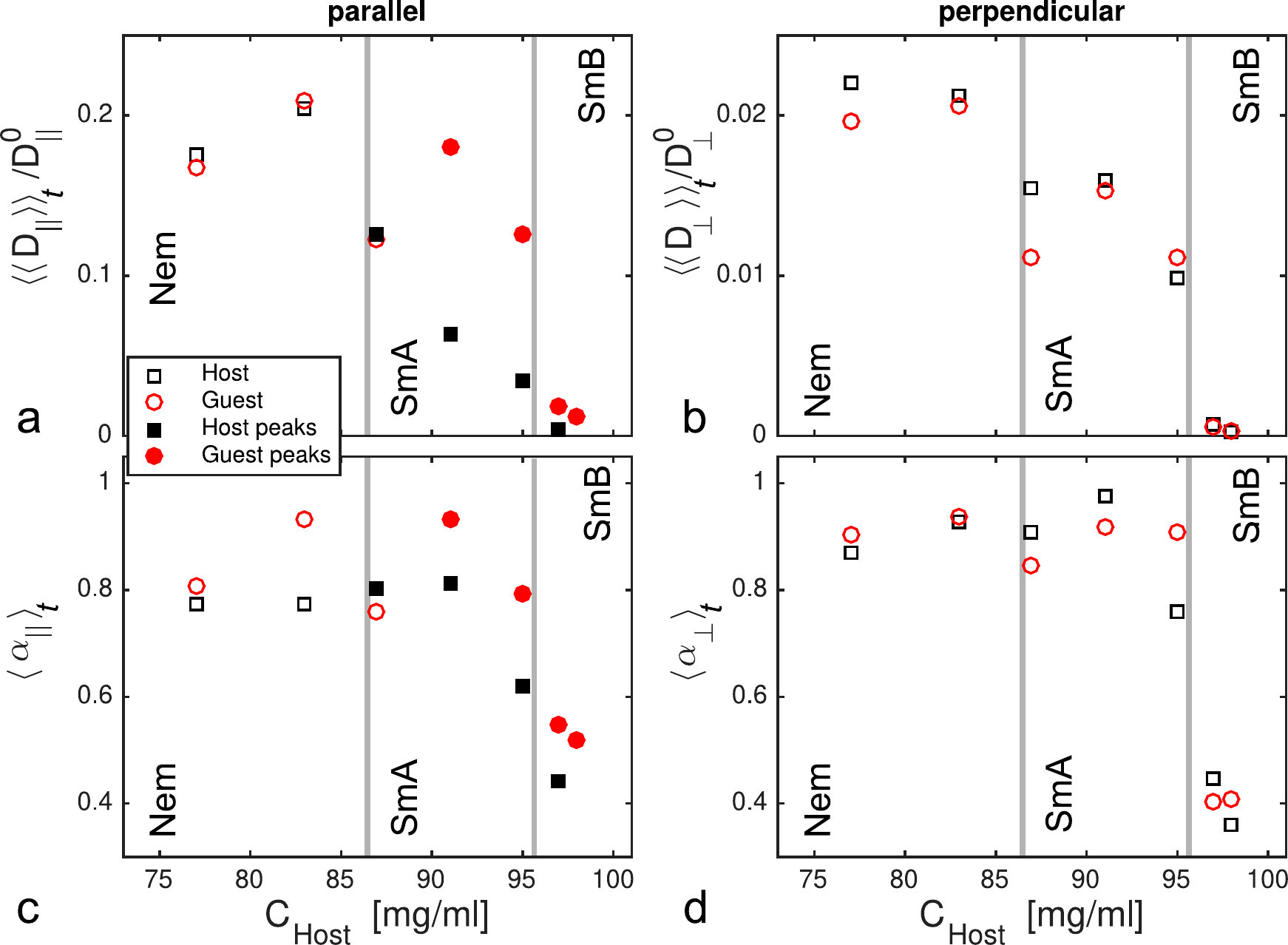}
	\end{center}
		\caption{Time averaged values of $\langle D_{\|,\bot} \rangle$ (a,b) and $\alpha_{\|,\bot}$ (c,d) for parallel (a,c) and perpendicular (b,d) diffusion. The full symbols indicate the results obtained by fitting the envelope of the SvH at the peak values.  } 
		\label{Fig_svhresavr}
\end{figure}

\section{Discussion}

The strength of the SvH analysis is two fold. First, this is the most sensitive metrics to characterize the dynamics of the system, which is reflected in the distinct line shape of $G(r,t)$. Second, it provides an instantaneous measure of the dynamics at hand, in contrast to the $\gamma$ parameter from the MSD, which requires at least a decade in time for a proper determination. 
The fit of the SvH gives valuable information on the time-dependent dynamics of the system, which can be seen by plotting $\langle D \rangle$ and $\alpha$ as a function of time, as shown in Fig. \ref{Fig_svh_results} for the parallel and perpendicular directions. 
In the following, we will discuss the dynamic behavior in each mesophase of the phase diagram, which will give us insight of the distinct dynamics of both guest and host particles.

In the nematic phase, $\langle D_{\|} \rangle (t)$ and $\langle D_{\bot} \rangle (t)$ are relatively featureless, except that we do observe a slight initial decay both in $\langle D_{\|} \rangle (t)$ and $\langle D_{\bot} \rangle (t)$. 
The dynamics of both particles, as quantified by $\alpha (t)$ is close to one and does show some relaxation towards this value, especially in the perpendicular direction, as shown in Fig. \ref{Fig_svh_results}g and h.

When entering the smectic-A phase, we miss the initial time for the host particles, as it takes time to diffuse over two layers (See section above). This effect is more prononced as the dynamics is restricted by increasing the host particle concentration. 
The resulting $\langle D_{\|}^{host} \rangle (t)$ does not exhibit any time dependence, but it decreases continuously with increasing concentration, as does the exponent $\alpha_{\|}^{host} (t)$.
This shows that the energy landscape becomes more heterogeneous with concentration, due to the increase of the confining potential, as described earlier in Refs. \cite{Lettinga2007,Pouget2011,Alvarez2017}. Note, however, that the potential is merely a measure of the sharpness of the time-averaged  $\langle G(r_{\|}^{host},t)\rangle_t$, but it is not sensitive to the shape of $\langle G(r_{\|}^{host},t)\rangle_t$. 
The fact that $\alpha_{\|}^{host} (t)$ does not fully relax back to $\alpha_{\|}^{host}=1$ for long times suggests that the time of observation was not long enough for the rods to undergo many randomizing jumps, as required to recover effectively a Gaussian particle that undergoes random steps on a coarse grained time scale corresponding to the average time it needs between two hopping-type events.
Perpendicular diffusion of the host particles in the smectic-A phase, as quantified by $\langle D_{\bot}^{host}\rangle$ in Fig. \ref{Fig_svh_results}e, displays a slight initial decay with time accompanied by a slight increase of $\alpha_{\bot}^{host} (t)$, see Fig. \ref{Fig_svh_results}g. In this case, the long time limit seems to be reached. This dynamic behavior revealed by $\alpha_{\bot}^{host} (t)$ is typical for glassy behavior of colloidal spheres ~\cite{Kegel2000,Weeks2000,Weeks2002b} and polymeric glasses~\cite{Wang2012}.

The guest particles display a different behavior. For 87 mg/ml we still observe an initial decay in $\langle D_{\|}^{guest} \rangle (t)$, similar to the decay found in the nematic, while $\alpha_{\|}^{guest} (t)$ stays almost constant in the smectic-A phase, as the guest particles do not sense a strong potential. Further organization even seems to promote the dynamics, both in parallel and perpendicular direction. The relaxation of $\alpha_{\bot}^{guest} (t)$ is more pronounced for the guest particle as compared to the host particles. 
As the length of the guests does not fit to the length scale associated with the host surrounding phase, here the smectic layer spacing $L_{layer}$, the guest particles belong simultaneously to at least two adjacent smectic layers. Therefore, this creates transient voids within adjacent layers, which act as excluded volume for the host particles, see Fig. \ref{fig_guesthost}a, and as free volume for the guest ones. This free volume promotes the parallel self-diffusion of the latter and it is decoupled from the heterogeneous in-plane dynamics of the host particles. As relaxation needs to take place in two layers simultaneously, the relaxation of the perpendicular dynamics takes longer than for the host particles. As a result, the anisotropy in the diffusion of the guest particles in the smectic-A phase diverges, see Fig. \ref{Fig_diffani}.

When entering the smectic-B phase at 97 mg/ml, there is a very pronounced reduction of all dynamics of the guest particles. Apparently, the distorting effect of the non-commensurate particles does not affect the crystal structure of the smectic-B phase so that all dynamics is frozen.  
Accordingly, $\alpha_{\|}^{guest}$ drops to a value even slightly smaller than 0.5, indicating very sub-diffusive behavior, see Fig. \ref{Fig_svh_results}d. The same behavior is observed for the host particle, but  it is more pronounced for the guest particle, as the dynamics of the host particle already slows down throughout the smectic-A phase.

The difference between guest and host can be summarized by plotting the time-averaged values  $\langle \langle D_{\|} \rangle\rangle_t$ and $\langle\alpha_{\|}\rangle_t$ as a function of the concentration, as shown in Fig. \ref{Fig_svhresavr}a and c. Here we distinguish  between the values as obtained from fitting the full $G(r_{\|},t)$ (open symbols) or only at the peak positions (full symbols), as required when peaks are present, which is clearly at higher concentrations for the guest particles. The results for  $\langle \langle D_{\|}^{host,guest} \rangle\rangle_t$ are  in very good quantitative agreement to those obtained from the direct MSD analysis, see Fig. \ref{Fig_mdsres}a and \ref{Fig_svhresavr}a, confirming \textit{a posteriori} the choice of our fitting function in Eq. \ref{Eq_a1a2}. This is interesting, as the denominator in the exponent in Eq. \ref{Eq_a1a2} is corrected for the Gaussian limit, as it should, \textit{as well as} in the Laplacian limit. This is the more surprising as this factor is missing in Refs. \cite{Chechkin2017,Chubynsky2014}. 
Note also that the dynamics colloidal spheres in a periodic sinusoidal potential has been analysed by fitting the full peaked self-van Have function~\cite{Dalleferrier2011}. The fundamental difference with the rods in a periodic smectic potential is, however, that we cannot assume a static smooth sinusoidal potential.

The comparison of $\gamma_{\|}$ (Fig. \ref{Fig_mdsres}c) and $\langle\alpha_{\|}\rangle_t$ (Fig. \ref{Fig_svhresavr}c) reveals a marked difference. Where $\gamma_{\|}$ is basically constant with concentration for both particles, $\langle\alpha_{\|}^{host}\rangle_t$ shows a decay towards the smectic-A to smectic-B transition, while $\langle\alpha_{\|}^{guest}\rangle_t$ stays nearly constant. 
This information on the heterogeneity of the dynamics is hidden in the concentration dependence of $\gamma_{\|}^{host}$. 
Thus, the SvH is  more sensitive in picking up the dynamics of the system so that the distinction between the long guest and the short host diffusive dynamics is more obvious from the analysis of the SvH. This approach based on SvH functions is, however, highly demanding in terms of statistics.

\section{Conclusions}

The assumption that large particles always diffuse slower than small ones is not generally valid when the length scale associated with the energy landscape formed by self-assembled host particles is smaller than the length of the guest particle. We proved this effect by evidencing a promoted permeation of non-commensurate long guest rods through self-assembled smectic layers of shorter host particles, using a suitable system of filamentous bacteriophages \cite{Alvarez2017}. To  explain this phenomenon, one  should consider the relative free volume accessible for the guest and host particles. As non-commensurate long rods are always simultaneously present in at least two layers, they generate their own voids creating more free volume than host particles and facilitating their parallel displacement.

Here we elucidated the physics of the surprising anomalous behavior by analyzing not only the MSD, but also the Self-van Hove functions $G(r,t)$. 
The latter appear to be a very sensitive and powerful tool to distinguish between the dynamic behavior of long and short particles. 
Through this analysis, we show that the dynamics of the host particles becomes non-gaussian, and therefore heterogeneous,  after the nematic-smectic A phase transition, especially in the parallel direction  even though $\gamma_{\|}^{host} \approx 1$. In contrast, the non-commensurate guest particles still display Gaussian dynamics for the parallel motion of the rods along the director, up to the smectic-B phase, whilst the perpendicular dynamics shows a long-time relaxation towards Gaussian dynamics.
Finally, this relatively straightforward self-assembled system which displays this continuous transition from Laplacian to Gaussian dynamics could aid the development of more accurate diffusivity models.

\section*{acknowledgments}
L.A. acknowledges IdEx Bordeaux (France) for financial support. O.K. acknowledges the International Helmholtz Research School of Biophysics and Soft Matter for financial support.

\section*{Data Availability}
All data measured in this work and copies of the matlab codes  used for analysis are available upon request.

\clearpage
	
\end{document}